\documentclass[pre,aps,twocolumn,showpacs,preprintnumbers,superscriptaddress]{revtex4}

\usepackage{graphicx}
\usepackage{dcolumn}
\usepackage{bm}
\usepackage{amssymb}
\usepackage{pifont}
\usepackage{amsmath}
\usepackage{times}
\usepackage[nooneline]{subfigure}

\begin{document}
\title{Spatiotemporal structure of Lyapunov vectors in chaotic coupled-map lattices}

\author{Ivan G.\ Szendro} \email{szendro@ifca.unican.es}

\affiliation{Instituto de F\'{\i}sica de Cantabria (IFCA), CSIC--Universidad de
Cantabria, E-39005 Santander, Spain}

\affiliation{Departamento de F{\'\i}sica Moderna, Universidad de
Cantabria, Avda.~Los Castros, E-39005 Santander, Spain}

\author{Diego Paz\'o} 

\author{Miguel A. Rodr{\'\i}guez} 

\author{Juan M. L{\'o}pez} 

\affiliation{Instituto de F\'{\i}sica de Cantabria (IFCA), CSIC--Universidad de
Cantabria, E-39005 Santander, Spain}

\date{\today}

\begin{abstract}
The spatiotemporal dynamics of Lyapunov vectors (LVs) in spatially extended chaotic
systems is studied by means of coupled-map lattices. We determine intrinsic length scales
and spatiotemporal correlations of LVs corresponding to the leading unstable directions by
translating the problem to the language of scale-invariant growing surfaces. We find that
the so-called `characteristic' LVs exhibit spatial localization, strong clustering around
given spatiotemporal loci, and remarkable dynamic scaling properties of the corresponding
surfaces. In contrast, the commonly used backward LVs (obtained through Gram-Schmidt
orthogonalization) spread all over the system and do not exhibit dynamic scaling due to
artifacts in the dynamical correlations by construction.
\end{abstract}


\maketitle

Lyapunov exponents (LEs) measure the exponential separation (or convergence) of nearby
trajectories and provide the most common tool to characterize spatiotemporal chaos
(STC)~\cite{eckmann,Bohr,Ott}. Not only exponential separation rates but also spatial
correlations are crucial to deal with predictability questions in extended
systems~\cite{primo07}. Basically, (almost) any initial infinitesimal perturbation
evolves in time asymptotically aligning along the most unstable direction, {\it i.e.} the
main Lyapunov vector (LV). In extended systems, the main LV is found to be localized in
space at any given time. However, due to spatial homogeneity, all directions in tangent
space are actually equivalent and the localization center keeps moving all over the
system. This is known as `dynamic localization' of the main LV~\cite{pik94,pik98}.

In the last decade, there has been some progress in the study of STC with tools borrowed
from nonequilibrium statistical physics. In particular, the evolution of perturbations in
spatially extended chaotic systems has been shown to be described by multiplicative
Langevin-type equations~\cite{pik94,pik98,lopez,sanchez}. In many cases~\cite{pik98}, the
dynamics of perturbations can be expressed in terms of the prototypical stochastic surface
growth equation of Kardar-Parisi-Zhang (KPZ), $\partial_t h= (\partial_x h)^2 +
\partial_{xx} h + \zeta$, where $\zeta$ is a white noise~\cite{kpz}.

In this Rapid Communication we study the spatiotemporal structure of STC encoded by the
LVs. We show that a family of vectors ---that we shall call characteristic LVs--- carry
important information about the real-space structure, localization properties and
space-time correlations. These properties are disclosed  only after a logarithmic
transformation, so that each LV is mapped into a rough surface. Our results greatly
strengthen the link between STC and certain nonequilibrium surface growth models. We also
find that the most widely used orthogonal LVs ---that appear in the standard Gram-Schmidt
procedure to compute the LEs~\cite{benettin80,wolf85}--- lack many of these features due
to the construction procedure and, therefore, have much less physical significance.

We illustrate our results by numerical simulations of coupled map lattices (CMLs), which
are a prototype of spatially extended dynamical systems exhibiting chaos. The main
advantage of CMLs, as compared with {\it e.g.}, partial differential equations, is their
lower computational cost. We consider an array of $L$ maps diffusively coupled to
nearest-neighbors:
\begin{equation}
\label{cml}
u_i(t+1)= \epsilon f(u_{i+1}(t))+\epsilon
f(u_{i-1}(t))+(1 - 2\epsilon )f(u_i(t)),
\end{equation}
where $\epsilon$ is the coupling parameter and $f$ a map with chaotic dynamics.
Throughout this paper, we consider the logistic map $f(y)=4y (1-y)$ with periodic
boundary conditions. We follow the standard convention and sort LEs in decreasing order
$\lambda_1 \geq \lambda_2 \geq \cdots \geq \lambda_L$. The coupling parameter is chosen to
be $\epsilon=0.1$, and the system is hyperchaotic with $\lambda_n >0$ for $n/L < 0.795$.

{\em Lyapunov vectors.-} Given the initial state of the system ${\bm u(t=0) =
[u_1(0), u_2(0), \cdots, u_L(0)]}$, any infinitesimally small change ${\bm {\delta
u}}(t=0)$ in the initial condition evolves up to linear order ({\it i.e.}~in tangent
space) according to
\begin{eqnarray}
&\delta u_i(t+1) = \epsilon f^{\prime}(u_{i+1}(t))\delta u_{i+1}(t)
+\epsilon f^{\prime}(u_{i-1}(t))\delta u_{i-1}(t)& \nonumber\\ &+ (1
-2\epsilon)f^{\prime}(u_i(t))\delta u_i(t)
\equiv \sum_{j=1}^L T_{ij}[{\bm u}(t)]\delta u_j(t)&,
\label{tangent}
\end{eqnarray}
with $f^{\prime}$ being the derivative of $f(y)$ with respect to $y$ and
$\mathrm{\mathbf{T}}[{\bm u}(t)]$ the $L \times L$ Jacobian matrix evaluated at ${\bm
u}(t)$. The evolution of an infinitesimal perturbation $ \bm{\delta u}(t_1)$ is governed
by the linear equation (tangent space): ${\bm {\delta
u}}(t_2)=\mathrm{\mathbf{M}}(t_2,t_1) {\bm{\delta u}}(t_1)$. The linear operator
$\mathrm{\mathbf{M}}$ is just the product of the Jacobian matrices evaluated along the
system trajectory from $t_1$ to $t_2$, {\it i.e.} $\mathrm{\mathbf{M}} (t_2,t_1)\equiv
\mathrm{\mathbf{T}}[{\bm u}(t_2-1)]\mathrm{\mathbf{T}}[{\bm u}(t_2-2)]\ldots
\mathrm{\mathbf{T}}[{\bm u}(t_1)]$.

According to Oseledec's theorem~\cite{oseledec} (details can be found in~\cite{eckmann})
there exists the limit operator ${\bm \Phi}_{\infty}(t_2) = \lim_{t_1 \to -\infty}
[{\mathrm{\mathbf{M}}(t_2,t_1)} {\mathrm{\mathbf{M}}(t_2,t_1)}^*]^{\frac{1}{2(t_2-t_1)}}$,
such that the logarithms of the eigenvalues are the LEs $\lambda_n$, and the eigenvectors
form an orthonormal basis $\{{\mathbf{e}_n(t)}\}$. This set of eigenvectors, so-called
{\em backward} LVs~\cite{legras96}, indicates the directions of growth of perturbations
grown since the remote past. The backward LVs are precisely the orthonormal vectors
obtained using the standard Gram-Schmidt orthogonalization method to compute the
LEs~\cite{ershov98}.

As we will show below a much more interesting set of vectors $\{{\mathbf{g}_n(t)}\}$ are
the so-called {\em characteristic} LVs. These are free-evolving perturbations that grow
with exponent $\lambda_n$ ($- \lambda_n$) when integrating to the far future (past):
$\lim_{|t_2| \to \infty} (t_2-t_1)^{-1} \ln ||{\mathrm{\mathbf{M}}}(t_2,t_1) \mathbf{g}_n
(t_1) ||= \lambda_n$. Characteristic LVs are independent of the scalar product and do not
form an orthogonal basis, in contrast to backward LVs. Also, they exhibit remarkable
physical properties as they reduce to the Floquet eigenvectors for a periodic
orbit~\cite{trevisan98}; and for autonomous continuous-time systems the null LE is
associated with a characteristic LV tangent to the trajectory. This contrasts with the
counterintuitive arrangement of backward LVs. We want to emphasize that characteristic LVs
evolve freely, whereas backward LVs (other than the first one) do not map into themselves
under time evolution.

Recently, characteristic LVs are receiving growing
attention~\cite{legras96,trevisan98,wolfe_tellus07,ginelli07}. Unfortunately, these
vectors are difficult to compute because information from both the remote past (via
backward LVs), and the remote future (via forward LVs) is needed~\footnote{Forward LVs are
defined as backward ones, but integrating to the past in tangent space (this requires the
storage of a long fiducial trajectory.}. We have computed the characteristic LVs from the
intersection of embedded subspaces from backward and forward LVs as described in
Refs.~\cite{legras96,wolfe_tellus07}. The $n$th characteristic LV is a linear combination
of the backward LVs from 1 to $n$, and viceversa; and in particular ${\mathbf{g}_1(t)} =
{\mathbf{e}_1(t)}$. Our computational resources allowed us to achieve the very reasonable
system size of $L=1024$.
\begin{figure}
\centerline{\includegraphics *[width=75mm]{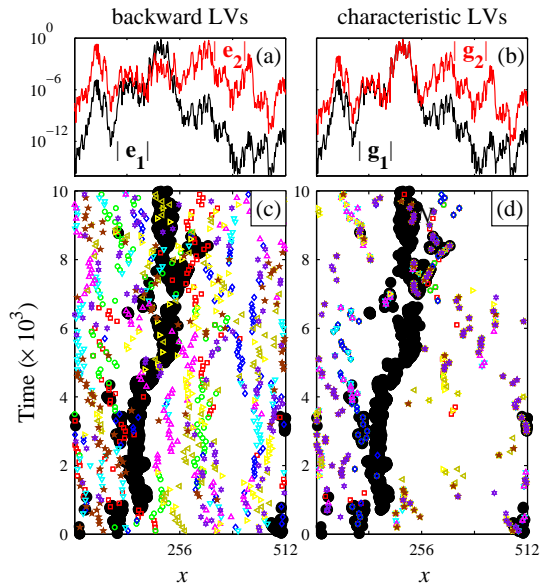}}\caption{(Color online) (a,b)
Snapshot of the profiles of LVs with $n=1,2$ for system~\eqref{cml} with $L=512$ (note that
we are taking the absolute values and a logarithmic scale). (c,d) Space-time evolution of the maxima
of LVs from the 1st to the 10th. The position of the maximum for the first LV is indicated
with big $\bullet$ symbols. Other symbols correspond to LVs for $n=2,\ldots,10$.
\label{fig1}}
\end{figure}

{\em Localization of Lyapunov vectors.-} Figure~\ref{fig1}(a,b) shows the first and
second LVs in logarithmic scale; the plots indicate strong localization. In
Fig.~\ref{fig1}(c,d) we plot the time evolution for the maxima of $|\mathbf{e}_n(t)|$ and
$|\mathbf{g}_n(t)|$ for $n=1,\ldots,10$ backward and characteristic LVs, respectively. We
see that maxima are spread all over the system in the case of backward LVs. However,
the localization positions of different characteristic LVs appear in clusters, so that
different vectors tend to be highly correlated, indicating that unstable manifolds are
nearly tangent. This clearly shows that the spread of localization positions in the case
of backward LVs is a byproduct of orthogonalization and does not represent properly the
spatial structure of the unstable directions. Extensivity of STC has often been
interpreted as the independent contribution of individual chaotic degrees of freedom
located at different positions of the system. This has been claimed to be supported by the
fact that backward LVs do localize at different positions~\cite{egolf00}. However, this
picture of STC in terms of almost independent building blocks, although appealing, becomes
problematic if we consider the characteristic LVs as the proper vectors truly encoding
the system dynamics.
\begin{figure}
\centerline{\includegraphics *[width=75mm]{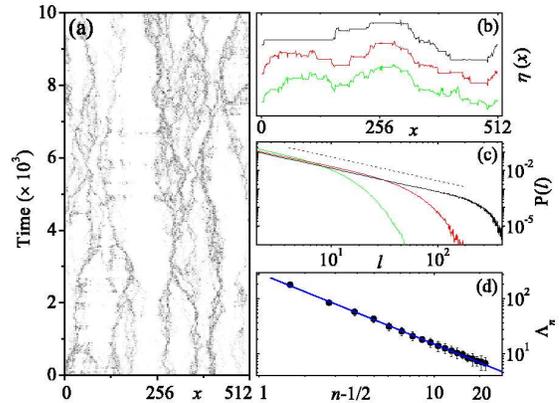}}
\caption{(Color online) (a) Spatiotemporal plot in gray scale of  $|\partial_x \eta_2|$
for the backward LVs. The plateaus of $\eta_2$ appear as white regions. (b) Snapshots of
$\eta_n, n=2,5,20$ from top to bottom (the curves are shifted to improve the visibility).
(c) Probability distribution of plateau lengths of the same $\eta_n$. The dashed straight
line is a power law $\sim l^{-1.28}$. (d) The cut-offs of the plateau distributions as a
function of $n$; the (blue) line is a guide to the eye with slope $-1.29$.\label{fig2}}
\end{figure}

{\em Surface growth picture.-} The scaling properties of the first LV were studied
by Pikovsky {\em et al.}~\cite{pik94,pik98}. For a wide class of extended dynamical
systems, they concluded that $\ln |\mathbf{e}_1(t)|$ can be interpreted as a growing
surface exhibiting dynamic scaling properties in the universality class of KPZ. In the
same spirit, here we define a surface for every (backward or characteristic) LV,
${\mathbf{f}_n(t)} \equiv [f_n(x,t)]_{x=1}^{x=L}$, via the Hopf-Cole transformation,
$h_n(x,t)=\ln |f_n(x,t)|$. Under this logarithmic transformation the $n$th LE maps into
the average velocity of the corresponding $n$th surface.

Some important features of the spatial structure are revealed by introducing the auxiliary
fields, $\eta_n \equiv h_n-h_1$, measuring the relative growth of the $n$th LV with
respect to the first one. In Fig.~\ref{fig2}(b) we plot snapshots of $\eta_n(x,t)$ for
backward LVs with $n=2, 5$ and $20$ (similar profiles are observed for characteristic
LVs). The profiles exhibit a distinctive structure organized in flat regions. These
plateaus evolve so that their boundaries have an erratic motion as can be seen in
Fig.~\ref{fig2}(a). This spatial structure results from the fact that the dynamics of
every LV is governed by Eq.~\eqref{tangent} with the same `deterministic noise'. Spatial
regions where $\partial_x\eta_n(x,t) \approx 0$ correspond to regions where the $n$th LV
surface closely follows the first LV. This means that the dynamics specific of the $n$th
LV surface is basically associated with that of the boundaries between plateaus. Since the
first LV surface dynamics belongs to the KPZ universality class, the picture that
immediately emerges from Fig.~\ref{fig2}(b) is that the $n$th LV surface is, loosely
speaking, ``piecewise KPZ".

The probability distributions of plateau lengths decay in good approximation as a power
law $\mathrm{P}(l) \propto l^{-\gamma}$ for $l< \Lambda_n$ with $\gamma \approx 1.28 \pm
0.10$ [Fig.~\ref{fig2}(c)]. We find good evidence that the cut-off scales as
$\Lambda_n\sim [L/(n-1/2)]^{\chi}$ where $\chi = 1.29 \pm 0.2$ [Fig.~\ref{fig2}(d)]. There
is some numerical uncertainty in the plateau and cut-off determination but we can estimate
the average plateau length $\langle l_n \rangle\approx [L/(n-1/2)]^{0.75 \pm 0.2}$.
\begin{figure}
\centerline{\includegraphics *[width=75mm]{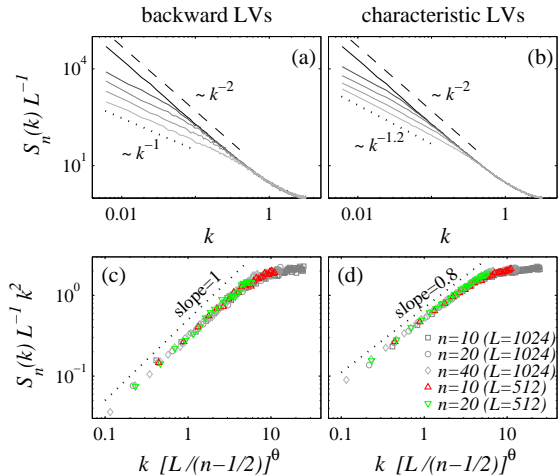}}\caption{(Color online) (a,b)
Structure factors of $h_n$, $n=1,5,10,20,40$ from top to bottom, $L=1024$. (c,d) Data
collapse through scaling relation~\eqref{supers} with fitting parameter $\theta=0.9 \pm
0.1$. \label{fig3}} 
\end{figure}

{\em Spatial structure.-} Let us now study the spatial correlations of the LV
surfaces $h_n(x,t)$. We computed the stationary structure factor $S_n(k)= \lim_{t \to
\infty} \langle \hat h_n(k,t) \hat h_n(-k,t) \rangle$, where $\hat h_n(k,t)= \sum_x
\exp(ikx) h_n(x,t)$. One expects the first LV surface correlations to decay as $k^ {-2}$
[Fig.~\ref{fig3}(a,b)], in accordance with KPZ universality~\cite{pik94,pik98}.
Interestingly, the $n$th LV surface for $n > 1$ also shows scale-invariant correlations
$\sim k^{-2}$, but only up to a crossover length scale that depends on $n$. At long
wavelengths correlations of LV surfaces, associated with backward and characteristic LVs,
decay as $k^{-1}$ and $k^{-1.2}$, respectively (Fig.~\ref{fig3}). This $1/k$-divergence
indicates extremely weak long-range spatial correlations for both classes of LVs.

Numerical data for all LV surface spatial correlations in the stationary state
[Fig.~\ref{fig3}(a,b)] can be cast in a single scaling function as shown in
Fig.~\ref{fig3}(c,d):
\begin{equation} S_n(k) L^{-1} k^2 = g(k/\Bbbk_n)
\label{supers}
\end{equation}
where $g(u)=\mathrm{const}$ for $u \gg 1$, and $g(u) \sim u^\sigma$ for $u \ll 1$,
$\sigma= 1$ and $\sigma=0.8$ for backward and characteristic LVs, respectively. The
crossover wavelength is $\Bbbk_n \sim [L/(n-1/2)]^{-\theta}$, with $\theta=0.9 \pm 0.1$,
and gives a typical length
scale $\ell_n \propto \Bbbk_n^{-1}$ for the $n$th LV surface, which immediately suggests
to link $\ell_n$ with the average plateau length $\langle l_n \rangle$ introduced
above.
\begin{figure}
\centerline{\includegraphics *[width=75mm]{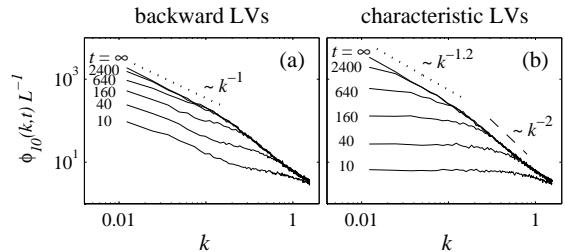}}
\caption{Time evolution of $\phi_{10}(k,t)$  for (a) backward and (b) characteristic LVs
($L=512$).\label{fig4}}
\end{figure}

{\em Space-time correlations.-} For each $n$ we consider the field
$h_n(x,t_0+t)-h_n(x,t_0)$, measuring the relative growth of the $n$th LV between times
$t_0$ to $t_0+t$ at a given point in space and whose structure factor we denote by
$\phi_n(k,t)$. We run trajectories for a long time $t_0$ to allow a good projection into
the attractor and then measure $\phi_n(k,t)$ at different times $t$. Figure~\ref{fig4}(a)
shows that backward LV surfaces for $n>1$ show long-range correlations $\sim 1/k$ already
at very short times. We claim this is due to non-local information introduced by the
Gram-Schmidt (orthogonalization) procedure that involves all LVs with $n'<n$.
Contrastingly, characteristic LV surfaces [Fig.~\ref{fig4}(b)] exhibit excellent dynamic
scaling properties. In this case the  $n$th surface is uncorrelated ---as indicated by a
flat $\sim k^0$ structure factor--- at scales $k \ll t^{-1/z}$, where $z$ is the dynamic
exponent. As time proceeds correlations develop and extend over a correlation length $\xi
\sim t^{1/z}$, as expected for typical dynamic scaling in kinetic surface roughening.
Eventually, saturation occurs when $\xi$ reaches the system size $L$, and we recover the
stationary correlations $\phi_n(k,t \rightarrow \infty) = S_n(k)$.

\begin{figure}
\centerline{\includegraphics *[width=85mm]{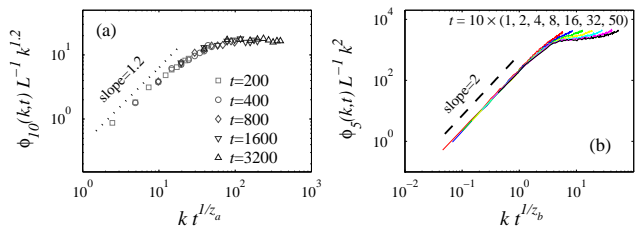}}\caption{(Color online) Dynamic
scaling through scaling relation~\eqref{dynamic}. (a) Data collapse above the crossover
with $\mu_a=1.2$ and $z_a=1$. Scaling shows up only for wavenumbers smaller than the
intrinsic scale $k < \Bbbk_{10} \approx 0.13$, in agreement with the crossover picture.
(b) Collapse below the crossover, $n=5$ (instead of $n=10$) has been used to obtain a
larger region of scaling; $\mu_b=2$ and $z_b=1.75 \pm 0.1$.\label{fig5}}
\end{figure}

In agreement with the results for the static correlations, $S_n(k)$, discussed above, we
can see two scaling regimes in Fig.~\ref{fig4}(b) separated by the typical wavenumber
$\Bbbk_n$. Curves corresponding to different times in Fig.~\ref{fig4}(b) can be
`collapsed' into:
\begin{equation}
\phi_n(k,t) = k^{-\mu_{i}} c_n^{(i)}(kt^{1/z_{i}})
\label{dynamic}
\end{equation}
where the label $i \in \{a,b\}$ denotes the scaling region, {\em above} or {\em below} the
crossover. On the one hand, for $t \gg \Bbbk_n^{-z_b}$, we find $\mu_a=1.2$
[Fig.~\ref{fig5}(a)]. Our simulations indicate that the dynamic exponent $z_a=1$ is
consistent with numerics on this range of scales for all $n$. This suggests a ballistic
motion of the plateau boundaries of $\eta_n$. This small value of $z_a$ implies that the
surface height correlates very rapidly once the correlation length reaches the intrinsic
length scale $\ell_n$. Thus, the larger $n$ the sooner $\phi_n(k,t)$ reaches the
asymptotic value. On the other hand, for $t \ll \Bbbk_n^{-z_b}$ we find $\mu_b=2$ [Fig.\
\ref{fig5}(b)]. Our simulations indicate ---supported also by the computation of the
surface growth exponent $\beta_b$ (not shown)--- a slow continuous increase of the dynamic
exponent with $n$ (for $n=1$, $z_{KPZ}=3/2$ for sufficiently large systems). Therefore, at
short scales characteristic LV surface correlations propagate slightly more slowly when we look
at larger $n$.

{\em Observations and Outlook.-} In our opinion characteristic LVs are more
fundamental for a theoretical understanding of STC than the more commonly used
(Gram-Schmidt) backward LVs. We have shown that they have strong localization
properties, so that they generically localize in physical space, signaling regions where
most unstable expanding directions concentrate at a given time. This suggests that
characteristic LVs, rather than backward vectors, may play an important role in the
hierarchical decomposition of STC~\cite{egolf00}.

We conjecture that the dynamics of the $n$th characteristic LV surface for $n > 1$ is
described by a KPZ {\em saddle} solution: given a certain realization of the
(deterministic) noise there should exist (measure zero) initial perturbations that evolve
toward states whose space-time surface correlations scale with critical exponents that are
not given by the standard KPZ universality class. This would correspond to the scaling
$\sim k^{-1.2}$ for long wavelengths, $k < \Bbbk_n$, in Fig.~\ref{fig3}(b).
While solutions whose basin of attraction is a set of measure zero are
irrelevant for stochastic equations, in deterministic chaos these may be fundamental
solutions for the complete understanding of sensitive dependence in high-dimensional
chaotic systems. In fact, it has recently been shown that the stochastic KPZ equation does
exhibit this type of saddle solutions for noise realizations with suitable (and admittedly
bizarre) correlations~\cite{canet_moore07}. We have shown that, in statistical terms, the
$n$th characteristic LV surface $h_n(x,t)$ differs from a genuine KPZ surface in an
amount given by a field $\eta_n(x,t)$. It is not difficult to convince oneself that the
step field corresponds to the nontrivial solutions ({\em i.e.} $\eta_n \neq
\mathrm{const.}$) of the stochastic equation:
\begin{equation}\label{diego_eq}
\partial_t \eta_n = (\partial_x \eta_n)^2 + \partial_{xx}\eta_n +
2(\partial_x h_1)(\partial_x\eta_n).
\end{equation}
In a quite natural way the term $\partial_x \eta_n$ on the right hand side leads to solutions of the
type shown in Fig.\ \ref{fig2}(b); plateaus of random spatial extent separated by
boundaries moving erratically. We believe that Eq.~\eqref{diego_eq} may indeed provide a
mathematical tool to address questions like real-space structure of the dynamics of
extended systems, space-time correlations in the propagation of disturbances, degree of
localization of unstable directions, and extensivity and decomposition of STC.

Finally, we have found identical behavior in other CML models. In fact, our conclusions
should be generic for LVs corresponding to LEs well above zero in systems whose first LV
surface belongs to the KPZ universality class~\cite{szendro_unpu}. (This certainly should
include all those systems reported in Ref.~\cite{pik98}.) Although more research is needed
to test the generality of our findings, in particular for continuous-time systems, our
preliminary results with a minimal stochastic model ($\partial_t w = \zeta w +
\partial_{xx}w$) confirm the universality of our results~\cite{szendro_unpu}.\\

Financial support from the Ministerio de Educaci\'on y Ciencia (Spain) under projects
FIS2006-12253-C06-04 and CGL2004-02652/CLI is acknowledged. DP acknowledges support by
MEC (Spain) through the Juan de la Cierva Programme.

\bibliographystyle{prsty}

\end{document}